\documentclass[final,3p,times,preprint]{elsarticle}
\usepackage{epstopdf}
\usepackage[colorlinks]{hyperref}
\usepackage{amsmath,array,amssymb}
\usepackage{graphicx}
\usepackage{caption}
\usepackage{subcaption}
\usepackage[numbers]{natbib}
\usepackage[T1]{fontenc}
\usepackage{nameref}
\usepackage{siunitx}
\usepackage{booktabs}
\usepackage{url}
\graphicspath{{./images/}}
\usepackage{color}

\journal{Nuclear Instruments and Methods in Physics Research Section A}

\begin{document}

\begin{frontmatter}

\title{A new technique for infrared scintillation measurements}
\author[1]{F. Chiossi}
\author[2]{K. Brylew}
\author[3]{A. F. Borghesani}
\author[1]{C. Braggio}
\author[1]{G. Carugno}
\author[2]{W. Drozdowski}
\author[1]{M. Guarise}

\address[1]{Dip. di Fisica e Astronomia and INFN, University of Padua, Via F. Marzolo 8, I-35131 Padova, Italy}
\address[2]{Institute of Physics, Faculty of Physics, Astronomy and Informatics, Nicolaus Copernicus University, Grudziadzka 5, 87-100 Torun, Poland}

\address[3]{CNISM unit and Dip. di Fisica e Astronomia, University of Padua, Via F. Marzolo 8, I-35131 Padova, Italy}

\address{}

\begin{abstract}
We propose a new technique to measure the infrared scintillation light yield of rare earth (RE) doped crystals by comparing it to near UV-visible scintillation of a calibrated Pr:(Lu$_{0.75}$Y$_{0.25}$)$_{3}$Al$_5$O$_{12}$ sample. As an example, we apply this technique to provide the light yield in visible and infrared range up to \SI{1700}{nm} of this crystal. 

\end{abstract}


\end{frontmatter}

\section{Introduction}
\label{intro}

A new concept, all-optical particle radiation detector has been very recently proposed based on the mechanism of upconversion in RE-doped materials ~\cite{noi_2risonanza}. This process, in which low energy incident optical radiation (infrared light - IR) is converted into higher energy emitted photons (visible light),  is efficiently accomplished by incorporating RE ions in inorganic matrices due to their $f$-electrons configurations. 
In fact, ground state absorption (level 0) allows the rare earth ions to reach a metastable intermediate state (level 1), characterized by relatively long lifetimes ($\approx$ms), then another photon delivered by a pump laser tuned to the transition $1 \rightarrow 2$ promotes the ion to a more energetic state (level 2). Radiative transition from this latter excited state back to the ground state is then observed by means of traditional detectors as photomultipliers (PMT) or photodiodes (PD).  
To date, this mechanism has been extensively applied for the development of lasers and optical devices~\cite{scheps1996upconversion, goldschmidt2015upconversion} but its applicability in the field of particle detection has not been deeply investigated.

The visible light yield (LY) of this novel device depends on the upconversion efficiency and on the number of RE-ions excited into the metastable level 1 per energy unit of the particle. This latter quantity can be estimated by studying the LY in the IR band which very few articles in literature are concerned with \cite{moses1998prospects}. Actually, there is little interest in the RE-doped crystals infrared scintillation for their long decay lifetimes and for the low quantum efficiency of PMT in this spectral region.

The aim of this work is to propose a method that allows us to tackle a systematic investigation of the IR LY in several different materials, composed of different matrices, dopants and concentrations. This method is based on the luminescence comparison with a reference Pr:(Lu$_{0.75}$Y$_{0.25}$)$_{3}$Al$_5$O$_{12}$ single crystal whose light yield in the near UV-visible is known. Moreover, this method is applied to this crystal, thereby yielding its IR LY. 

\section{Experimental setup}\label{pro}

The mixed lutetium-yttrium aluminum garnet sample has been grown via Czochralsky method at ITME, Warsaw and its preparation is described elsewhere~\cite{drozdowski33000}.
The interest in mixed Pr:LuYAG crystals is related to its much better performance in terms of light yield and energy resolution as compared to either Pr:LuAG or Pr:YAG. The $5\times 5\times 5$ mm$^3$ sample chosen for the present measurements has a reported light yield of \SI{27000}{ph/MeV} and a \SI{5.3}{\%} energy resolution in the UV-VIS range~\cite{drozdowski33000}.

To investigate its light yield in the IR band, the Pr:LuYAG sample is excited by X-rays generated by an electron gun that can be operated both in continuous and in pulsed mode sweeping the electron beam at \SI{100}{Hz} frequency~\cite{barcellan2011battery}.   
The several \si{\micro\A} intense electron beam impinges on a $\sim$\SI{10}{\micro\m}-thick tantalum foil placed in front of the sample to make sure that the whole crystal is excited only by X-rays. The scintillation response of the crystal sample is measured using Si (mod. Hamamtsu S1337-1010BQ) and InGaAs (mod. Thorlabs DET20C) PDs. The small contribution to the PD signal of the X-rays that are not absorbed in the crystal and reach the PD can be estimated and then subtracted by covering the PD with an thin aluminum foil. 

In order to verify that the radioluminescence is determined by the Pr$^{3+}$ ions emission, the X-rays excitation spectra are compared with those obtained when the crystal is irradiated with a pulsed, frequency-quadrupled Nd:YAG laser (\SI{266}{nm}). In fact, whereas the host matrix is transparent at this wavelength, the Pr$^{3+}$ ions are directly excited into $4f5d$ levels, thus allowing us to simulate the energy transfer process from the electron-hole recombination to the RE ions after the X-ray excitation.

\section{Spectroscopic analysis}\label{spectroscopy}
\begin{figure*}[th]
\centering
\includegraphics[width=0.45\textwidth, angle =-90]{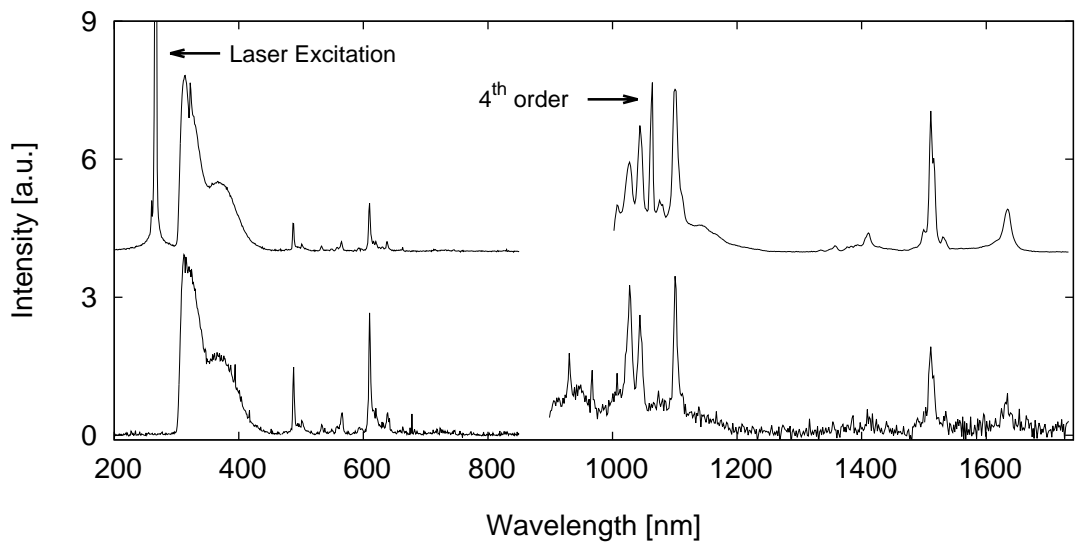}
\captionof{figure}{Uncorrected emission spectra of Pr:LuYAG under \SI{266}{nm} laser excitation \emph{(top)} and X-ray excitation \emph{(bottom)}.}
\label{fig:egun}
\end{figure*}

We have recorded spectra of the Pr:LuYAG emission due to X-ray and UV excitation (Fig.\,\ref{fig:egun}). The OceanOptics 650 RedTide and OceanOptics NIRQuest512 spectrometers were used for the 200--\SI{850}{nm} and 900--\SI{1700}{nm} regions, respectively.

In both spectra one can clearly see emissions related to the same Pr$^{3+}$ energy levels. We observe a strong  $4f5d \rightarrow 4f^2$ emission in the range from 300 to \SI{450}{nm} and narrow lines in the visible/near-infrared region due to emission from levels $^{3}\mathrm{P}_0$, $^{1}\mathrm{D}_2$, $^{1}\mathrm{G}_4$. The spectra below \SI{850}{nm} are similar to those reported by previous authors in a Pr:LuAG crystal~\cite{drozdowski2009effect}.
The main near-infrared transitions that we identify are: $^1\mathrm{D}_2 \rightarrow$ $^{3}\mathrm{F}_{3,4}$, $^1\mathrm{G}_4 \rightarrow$ $^3\mathrm{H}_4$ (900--\SI{1100}{nm}) and $^1\mathrm{D}_2 \rightarrow$ $^1\mathrm{G}_4$ (1400--\SI{1600}{nm}). 
The emission from $^1\mathrm{G}_4$ is expected to be partially non radiatively quenched, whereas the lower lying levels are strongly quenched in LuYAG matrix. As a consequence no mid-infrared emission is expected from these latter levels.  

\section{Method}

\begin{figure*}
\centering
\includegraphics[width=.33\textwidth, angle =-90]{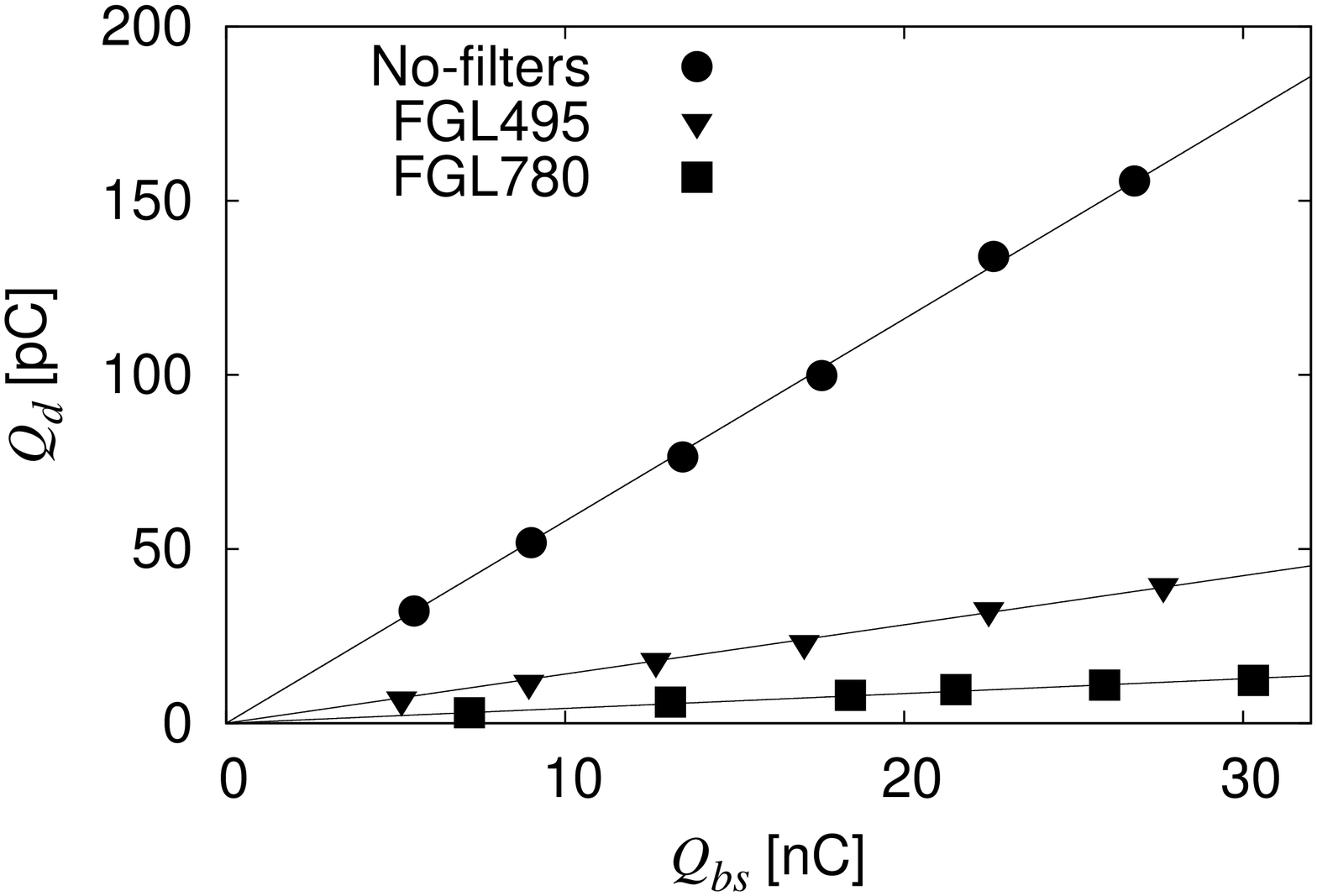}
\includegraphics[width=.33\textwidth, angle =-90]{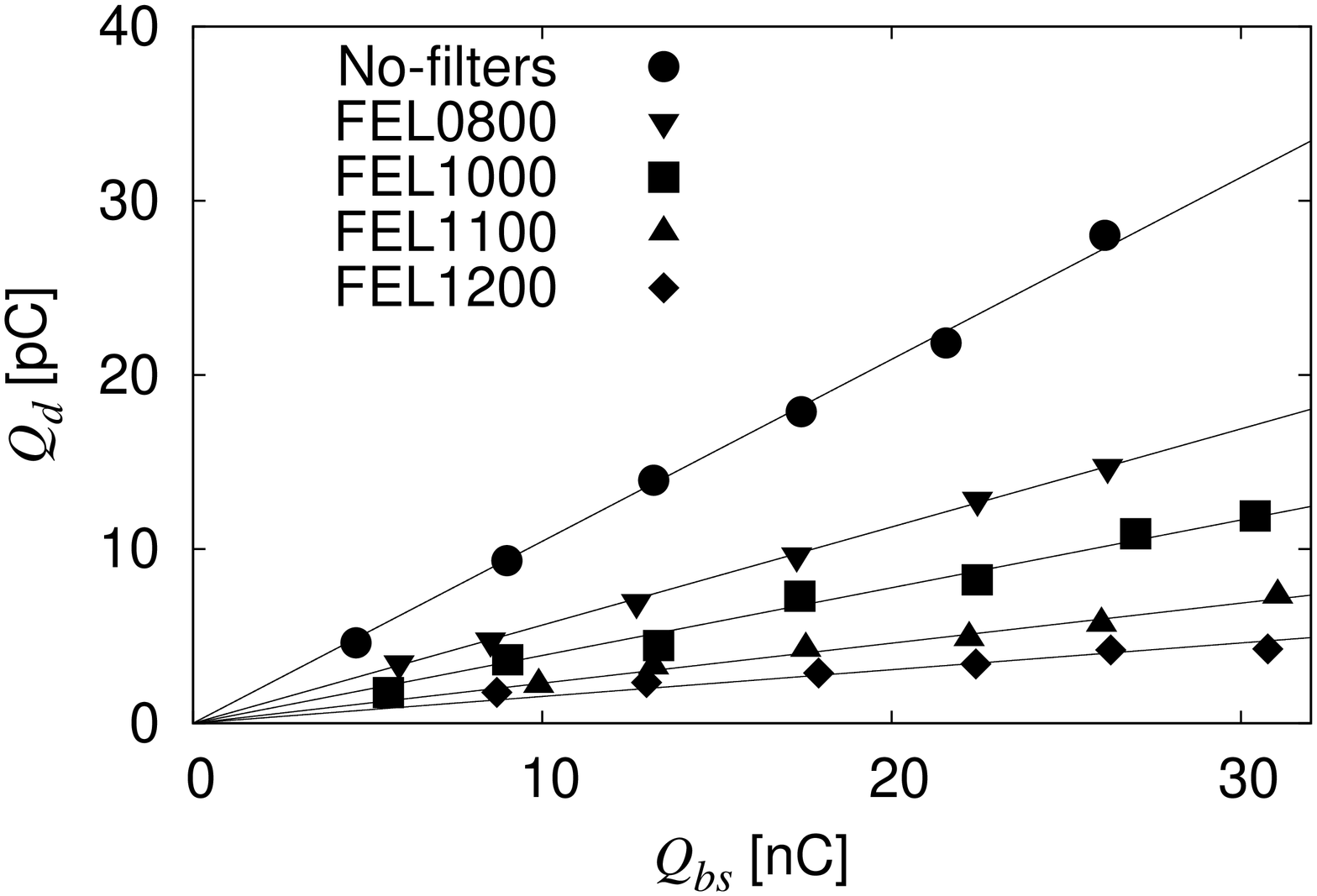}
\captionof{figure}{Linear dependence of the visible and infrared integrated luminescence signal on the charge collected at the Farday cup. Different sets of longpass filters have been used to obtain the Si PD \emph{(left)} data and the InGaAs \emph{(right)} data. The filter name contains the cut in wavelength in nm.}
\label{linearity}
\end{figure*}

We measure the total number of charge carriers generated per X-ray pulse in the photodiode $n_e=Q_d R_0/G\tau$ with $Q_d$ being the time integrated photodiode signal, $R_0 = $ \SI{1}{\mega\ohm} the impedance of oscilloscope, $G = \SI{0.25}{mV/fC}$  and $\tau \approx$ \SI{480}{\micro s} which are the gain and the time constant of the active integrator, respectively. If the PD quantum efficiency $\eta$ is constant in the considered range, $n_e$ is also given by equation  
\begin{equation}
n_e = E_{in} \cdot \mathrm{LY} \cdot \eta \cdot \dfrac{\Delta\Omega}{4\pi} \left( 1- R \right)
\label{eqn:main}
\end{equation}
where $E_{in}$ is the energy deposited in the sample, $R$ the crystal reflectivity, ${\Delta \Omega}=A/ d^2$ is the solid angle subtended by the PD with area $A$ located at a distance $d$ from the crystal. 
Moreover, the measured charge per pulse $Q_{bs}$ is related through a proportionality constant $k$ to the energy released by X-ray pulse in the crystal. In fact, as shown in the Fig.\,\ref{linearity}, the measured luminescence $Q_d$ linearly depends on $Q_{bs}$. 

\begin{figure*}
\centering
\includegraphics[width=0.33\textwidth, angle =-90]{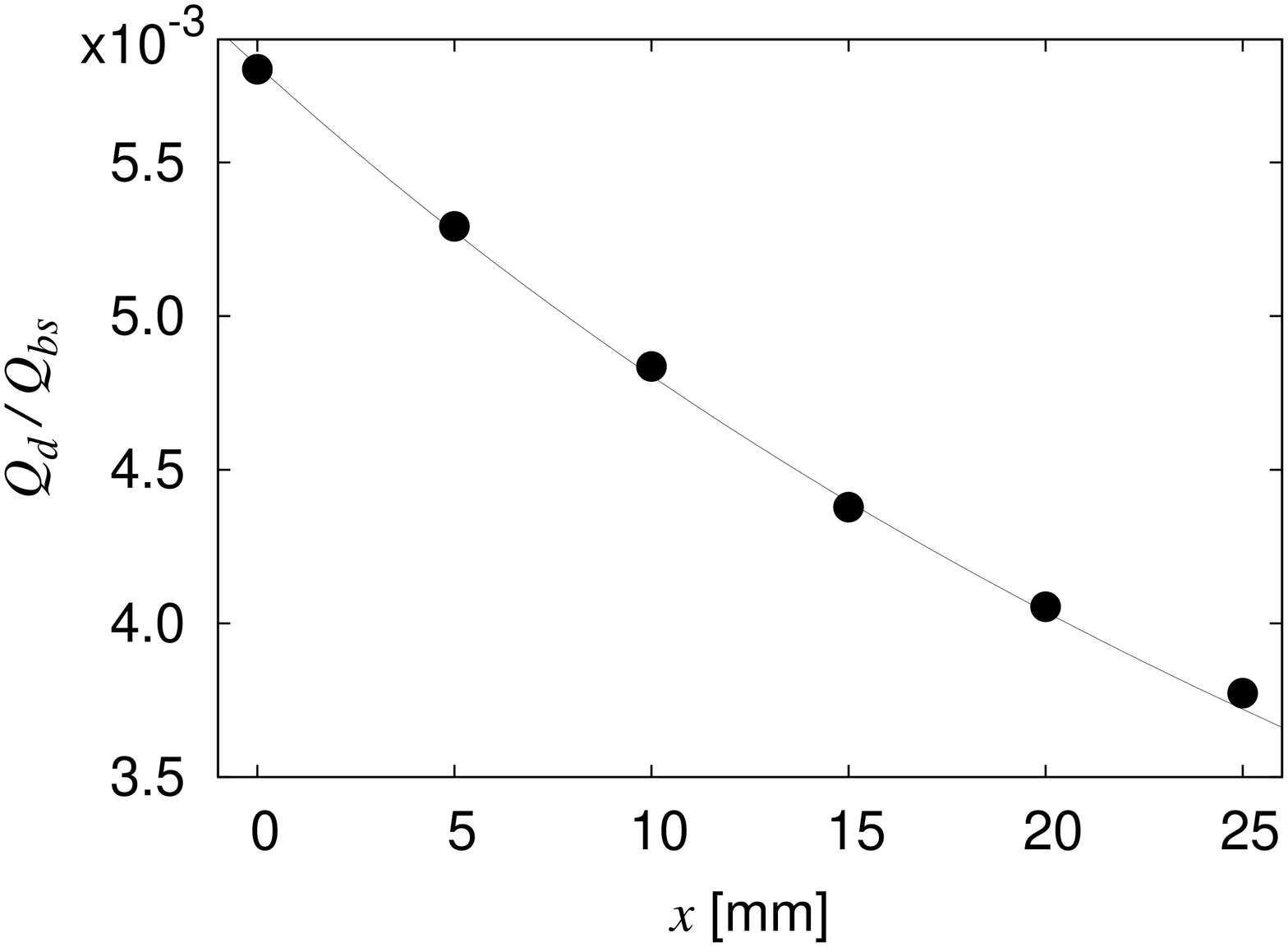}
\includegraphics[width=0.33\textwidth, angle =-90]{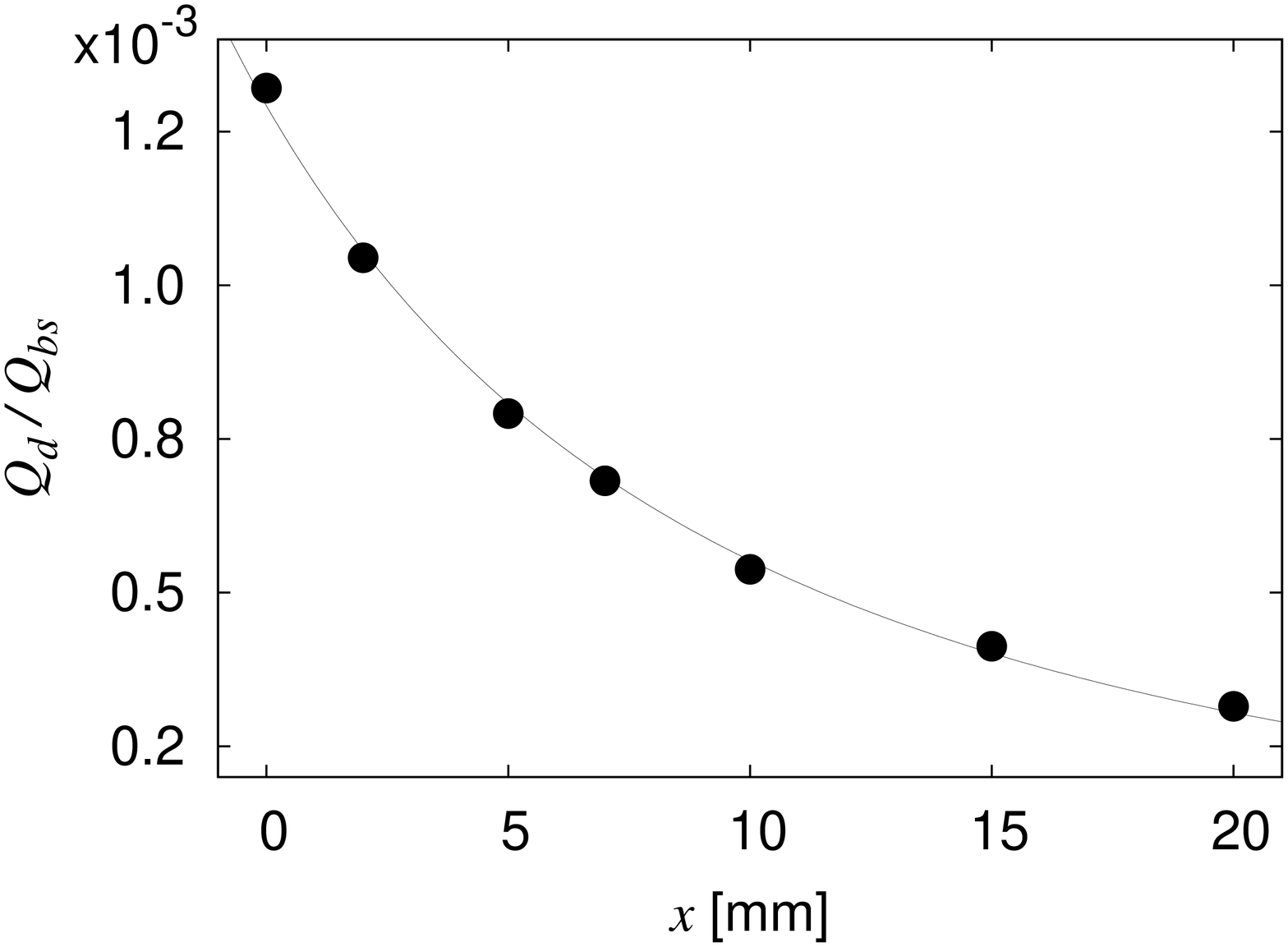}
\captionof{figure}{Measurement of the $a$ parameter with the Silicon \emph{(left)} and the InGaAs \emph{(right)} PD. The error bars
are of the same size as the symbols. The validity of the point source approximation is confirmed by the goodness of the fit.}
\label{poso}
\end{figure*}

It is then possible to obtain a general expression for the measured light yield in a definite wavelength range:
\begin{equation}
\mathrm{LY} =  \dfrac{Q_d}{Q_{bs}}d^2 \dfrac{4\pi R_0}{kG\tau} \dfrac{1}{\eta \left( 1-R \right)A}.
\label{eqn:almostLY}
\end{equation}
As in the point source approximation the following expression holds true
\begin{equation}
\dfrac{Q_{d}(x)}{Q_{bs}}= \left (\dfrac{Q_d}{Q_{bs}}d^2\right)  \dfrac{1}{\left( x_0 + x \right)^2} = \dfrac{a}{\left( x_0 + x \right)^2}
\label{eqn:sqr_r}
\end{equation}
the parameters $a=\dfrac{Q_d}{Q_{bs}}d^2$ and $x_0$ can be obtained by a fit of data recorded at different relative distances $x$ of the photodiode from the scintillating crystal, as shown in Fig.\,\ref{poso}.

We observe that in this type of measurements it crucial to precisely know the efficiency of the X-ray generation process, related to the previously mentioned proportionality constant $k$.  The latter can be estimated if the X-ray energy is fully released in the sample and provided the LY and the $a$ parameter of a generic crystal in any wavelength range are known.
As the sensitivity of the detector that has been used to measure the light yield of our LuYAG:Pr crystal reported in~\cite{drozdowski33000} peaks in the UV range, it is reasonable to expect that the reported light yield is mainly determined by $4f5d \rightarrow 4f^2$ radiative transitions. In addition, due to fast integration time the slow components might have been underestimated.  

 In order to select the photodiode signal component due to the UV scintillation, the previously described measurements are repeated at a fixed distance with optical longpass filters (Thorlabs, FGL and FEL filter sets) located in front of the photodiodes. 
 The use of filters allows us to estimate the light yield in narrower bandwidths and with a better accuracy by taking into account the wavelength dependence of the photodiode's responsivity. 

As the energy of the X-rays can be assumed to be in the range of few tens of keV, the LY of our sample measured in \cite{drozdowski33000} at \SI{662}{keV} has to be reduced by 10\%, in agreement with previously published data \cite{drozdowski2008scintillation, swiderski2009light, chewpraditkul2012comparison}. We report in Table \ref{tab:res} the results of the light yield measurements in several optical ranges.

\begin{table}[!htp]
	\centering
	\caption{Measured light yield of the LuYAG:Pr crystal for different optical ranges.}
	\begin{tabular}{l| c}
 	 \toprule
		Optical Band [nm]	& LY [ph/MeV]\\ 
	 \midrule
	 200--495 			& 24300 \\ 
	 495--780			& 6700 	 \\ 
	 780--1000			& 1000	 \\
	 1000--1100			& 1300 \\
 	 1100--1200			& 500\\  
	 1200--1700			& 1100\\  
	 \bottomrule 
	\end{tabular}  
\label{tab:res}
\end{table}

\section{Conclusions}\label{conclusions}

We have demonstrated a practical way to accurately estimate the infrared light yield of a RE-doped crystal by comparison with the near UV-visible luminescence of a reference crystal.  
The presented method allows us to make accurate LY measurements since it is based on the point-source approximation, as verified by performing measurements at several distances from the source, and because it is possible to vary the energy released per pulse in the crystal and to average over several measurements. Furthermore, the use of several optical longpass filters reduces the error due to the wavelength dependence of the photodiode quantum efficiency.

The method has been applied to a (Lu$_{0.75}$Y$_{0.25}$)$_{3}$Al$_5$O$_{12}$ crystal whose emission in the UV range had been previously measured. In spite of its high LY in the UV, its emission in the near infrared band is limited to a few thousands of ph/MeV, probably due to the $^1\mathrm{G}_{4}$ manifold quenching. Although the states of our interest $^3\mathrm{H}_\mathrm{J}$ and $^3\mathrm{F}_\mathrm{J}$, characterized by ms-long radiative lifetimes, are efficiently populated by the relaxation of the higher manifolds, no mid-infrared emission is expected from them in this host matrix.  

The data reported in this work can be used to extend the LY measurements up to the mid-infrared band, provided that photodiodes with a lower band gap than InGaAs are used.

In order to identify the most suitable crystals for the development of the upconversion-based detector, several RE-doped crystals are currently being investigated with the method described in this work. The preliminary results obtained for Nd:YAG and Tm:YAG are particularly encouraging for our aims and will soon be published.

\section{Acknowledgments}\label{thanks}
The growth of the crystals used in this research has been financed from the funds of the Polish National Science Centre granted on the basis of Decision no. DEC-2012/05/B/ST5/00324.

The authors FC, AFB, CB, GC, and MG acknowledge the technical assistance of Mr. E. Berto and Mr. L. Barcellan. 
	
  \bibliographystyle{elsarticle-num}

%%%%

\end{document}